\documentclass[conference]{IEEEtran}
\IEEEoverridecommandlockouts

\usepackage{cite}
\usepackage{amsmath,amssymb,amsfonts}
\usepackage{algorithmic}
\usepackage{textcomp}
\usepackage[table]{xcolor} 
\usepackage{multirow}
\usepackage{booktabs}        
\usepackage{multirow}        
\usepackage{graphicx}        
\usepackage{arydshln}        
\usepackage{textcomp}        
\usepackage{array}         
\usepackage{url}           
\usepackage{hyperref}
\def\BibTeX{{\rm B\kern-.05em{\sc i\kern-.025em b}\kern-.08em
    T\kern-.1667em\lower.7ex\hbox{E}\kern-.125emX}}
\usepackage{pifont}
\usepackage{amsmath,amssymb,amsfonts} 
\usepackage{mathtools} 
\usepackage[normalem]{ulem}
\useunder{\uline}{\ul}{}

\newcommand{\KL}{\mathbb{D}_{KL}} 
\newcommand{\cmark}{\ding{51}}%
\newcommand{\xmark}{\ding{55}}%
\begin{document}

\title{Omni-R1: Do You Really Need Audio to Fine-Tune Your Audio LLM?\\

}

\author{
Andrew Rouditchenko$^{1}$, Saurabhchand Bhati$^{1,*}$, Edson Araujo$^{2,*}$, \\
Samuel Thomas$^{3,4}$, Hilde Kuehne$^{2,4,5}$, Rogerio Feris$^{3,4}$, James Glass$^1$ \\
    \small\textit{$^1$MIT CSAIL \quad $^2$Goethe University of Frankfurt} \\
    \small\textit{$^3$IBM Research AI \quad  $^4$MIT-IBM Watson AI Lab \quad $^5$Tuebingen AI Center/University of Tuebingen} \\
    \small\texttt{roudi@mit.edu}
    \thanks{*Equal Contribution.}
}

\maketitle

\begin{abstract}
We propose Omni-R1 which fine-tunes a recent multi-modal LLM, Qwen2.5-Omni, on an audio question answering dataset with the reinforcement learning method GRPO.
This leads to new State-of-the-Art performance on the recent MMAU and MMAR benchmarks.
On MMAU, Omni-R1 achieves the highest accuracies on the sounds, music, speech, and overall average categories, both on the Test-mini and Test-full splits.
To understand the performance improvement, we tested models both with and without audio and found that much of the performance improvement from GRPO could be attributed to better text-based reasoning.
We also made a surprising discovery that fine-tuning without audio on a text-only dataset was effective at improving the audio-based performance.
\end{abstract}


\section{Introduction}

Reinforcement Learning (RL) has recently been shown to improve the reasoning capabilities of Large Language Models (LLMs)~\cite{guo2025deepseek}.
We are motivated by these advancements to improve the capabilities of Audio LLMs - models which take in audio input and text and can perform tasks such as question answering.
Building on Qwen2.5-Omni~\cite{qwen-omni}, a State-of-the-Art (SOTA) multi-modal LLM, we introduce Omni-R1 using a fine-tuning pipeline based on the RL method Group Relative Policy Optimization (GRPO)~\cite{shao2024deepseekmath}. 
Using a simple prompt, our model forgoes complex chain-of-thought or structured reasoning outputs and instead directly outputs answer choices.
We use the recent MMAU~\cite{sakshi2025mmau} and MMAR~\cite{ma2025mmar} benchmarks to test our models.
Fine-tuning on the audio and human-annotated questions from the AVQA~\cite{yang2022avqa} dataset boosts Qwen2.5-Omni’s average accuracy on MMAU Test-mini from 65.9\% to 68.6\%, and on the Test-full split from 68.4\% to 70.8\%. 

We also propose to automatically generate audio question answering datasets.
We prompt ChatGPT with several audio captions from an audio LLM to generate questions and answer choices for the 40k audios in AVQA and 182k audios in VGGSound.
Scaling up the training data with our automatically generated questions results in further gains, achieving new SOTA of 71.3\% on Test-mini and 71.2\% on Test full.
We obtained similar gains on the MMAR~\cite{ma2025mmar} benchmark leading to SOTA results.

In probing where these improvements arise, we made two surprising discoveries. 
First, text-only fine-tuning - dropping all audio inputs and training solely on question–answer text - results in significant gains in audio performance (e.g., Qwen2.5-Omni’s Test-mini average improves from 65.9\% to 68.2\% after fine-tuning on text-only science questions). 
Second, evaluation with audio withheld at inference shows that much of the performance boost originates from improved text-based reasoning rather than enhanced audio processing.

In summary, our contributions are:
\begin{itemize}
    \item Omni-R1, a streamlined GRPO fine-tuning on Qwen2.5-Omni that achieves new SOTA on MMAU and MMAR without any complex prompts or explicit reasoning.
    \item Automatically generated audio question answering datasets which scale question–answer pairs across 182k VGGSound clips to further boost performance.
    \item Analysis of text-only fine-tuning, demonstrating that improving an audio LLM’s text reasoning yields larger-than-expected gains on audio benchmarks.
\end{itemize}

We plan to release our code, models, and datasets publicly at \url{https://github.com/roudimit/Omni-R1}.

\section{Related Work}

Several recent large-scale audio question answering datasets were proposed to test audio LLMs, including MMAU~\cite{sakshi2025mmau} and MMAR~\cite{ma2025mmar}.
The MMAU benchmark~\cite{sakshi2025mmau} is a fixed answer choice dataset providing one correct answer and three incorrect answers for each question.
It contains questions about sounds, speech, and music of varying difficulty.
Some of the questions require external world-based knowledge that is not provided in the questions themselves.
MMAR~\cite{ma2025mmar} is a more recent benchmark designed to elicit deep reasoning from audio LLMs.
It includes questions about sounds, speech, and music, and also mixtures of them (for example, audio natively containing both sounds and speech).
The question and answer format is the same as MMAU.

GRPO is a Reinforcement Learning (RL) method proposed for instilling better reasoning capabilities in LLMs~\cite{shao2024deepseekmath,guo2025deepseek}.
Towards better Audio LLMs, R1-AQA~\cite{r1-aqa} proposes to use GRPO by fine-tuning Qwen2-Audio~\cite{qwen2-audio} on the AVQA~\cite{yang2022avqa} dataset of audio question and answers.
This method achieved the previous SOTA on MMAU, and our approach is inspired by them.
Using GRPO, we fine-tune Qwen2.5-Omni-7B~\cite{qwen-omni}, a recent multi-modal LLM proposed to handle both audio and video inputs, on AVQA.
Doing so, we were able to achieve the new SOTA on MMAU and also demonstrate strong performance on the MMAR benchmark, achieving SOTA results among open-source models and competitive performance to significantly bigger, close-ended models.
Further, we propose to automatically generate audio question and answer training data, which leads to even better performance.

SARI~\cite{sari} is a concurrent method which fine-tunes Qwen2.5-Omni using RL.
However, their setup is more complex than ours: they fine-tune the model with a schedule of Supervised Fine-Tuning (SFT) and RL, and use both structured reasoning and unstructured reasoning.
In contrast, we have a much simpler pipeline: we only fine-tune the model with RL, and we don't use any explicit reasoning.

Finally, we investigated how GRPO improves the Audio LLMs' performance by testing them with text-only inputs.
While recent work already tested audio LLMs with text-only inputs\cite{sakshi2025mmau,zang2025you}, we go a step further and fine-tune the models without audio (just text).
We made a surprising discovery that fine-tuning the models with text-only inputs could work almost as well as fine-tuning with audio.
It explains that much of the performance improvement from GRPO is due to improving the text-only reasoning capabilities.

\begin{figure}[t]
    \centering
    \includegraphics[width=\linewidth]{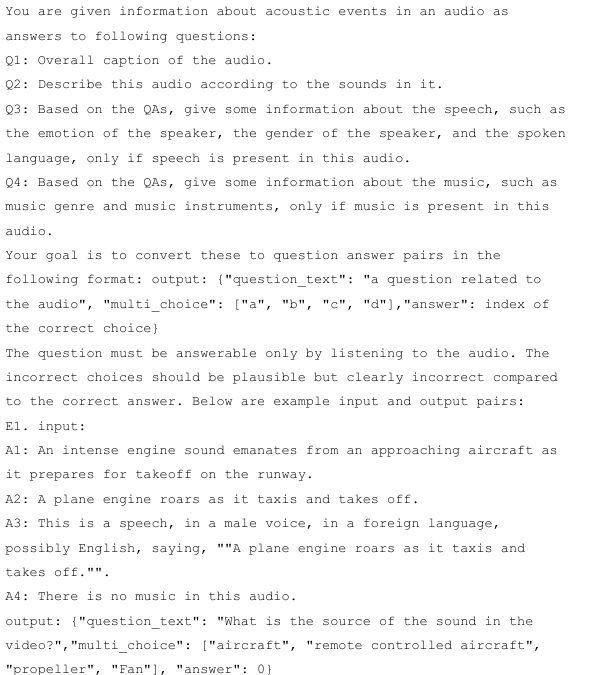}
    \caption{The prompt used with ChatGPT to generate our AVQA-GPT and VGGS-GPT datasets (Section~\ref{sec:dataset}).}
    \label{fig:overview}
\end{figure}

\section{Method}

\subsection{Audio LLM and Prompt}
We propose Omni-R1 which fine-tunes Qwen2.5-Omni-7B~\cite{qwen-omni} using GRPO.
During fine-tuning and inference, we use the prompt: \textit{``\texttt{<question>} Please choose the answer from the following options: \texttt{<choices>}''}.
We chose this prompt since it is simple and the model learns to directly output one of the answer choices.
This is especially useful since we have limited GPU memory.
Thanks to this simple prompt, we were able to perform full-finetuning on GPUs with only 48GB GPU memory, in contrast to recent works which perform fine-tuning on GPUs with 80GB GPU memory~\cite{r1-aqa}.
Although recent works~\cite{audio-reasoner,r1-aqa,sari} propose more complex prompts with reasoning, our simple prompt was sufficient for our models to achieve the new SOTA on MMAU and MMAR.

\subsection{Reinforcement Learning}
\label{sec:grpo}
Our training uses Group Relative Policy Optimization (GRPO)~\cite{r1-aqa, shao2024deepseekmath}, an adaptation of Proximal Policy Optimization (PPO)~\cite{schulman2017proximal}. GRPO is designed to mitigate the substantial memory overhead and the challenges of training an accurate per-token value function, common in standard PPO implementations for LLMs. GRPO achieves this by eliminating the explicit value function. Instead, it derives advantage estimates by comparing outputs within a group generated for the same input prompt, effectively using the average reward of sampled responses as a baseline.

The GRPO process begins by sampling $G$ distinct outputs, $\{o_1, o_2, \ldots, o_G\}$, for a given input question $q$ using a recent policy version, $\pi_{\theta_{old}}$. Each output $o_i$ is assigned a scalar reward $r_i$ by a reward model $r_\phi$. For outcome-based supervision, these rewards $\mathbf{r}=\{r_1, \ldots, r_G\}$ are normalized across the group to define the advantage. Specifically, for an output $o_i$, the advantage $\hat{A}_{i,t}$ is computed as:
\begin{equation} \label{eq:GRPO_advantage_revised}
    \hat{A}_{i, t} = \widetilde{r}_i = \frac{r_i - \operatorname{mean}(\mathbf{r})}{\operatorname{std}(\mathbf{r})},
\end{equation}
meaning this same advantage $\widetilde{r}_i$ is applied to all tokens $t$ within that specific output $o_i$.

The specific reward $r_i$ used in Equation~\eqref{eq:GRPO_advantage_revised} is derived from a rule-based reward function that evaluates model responses based on correctness: 
\begin{equation}
r_i = r_{acc}
\end{equation}
$r_{acc}$ is an accuracy reward: it is 1 if the model's response contains the correct answer, and 0 otherwise. 
We do not include any other rewards such as formatting rewards.

\begin{table*}[t]
    \renewcommand{\arraystretch}{1.3}
    \small
    \centering
    \caption{Accuracies (\%) on MMAU (original).
    SARI\textsuperscript{\textdagger} calculates scores using an LLM as a judge instead of the usual string matching.}
    \vspace{-2mm}
    \resizebox{0.88\linewidth}{!}{%
        \begin{tabular}{cccccc|cccc}
             \hline
            \toprule
             \multicolumn{1}{c}{\textbf{Model}} & \multicolumn{1}{c}{\textbf{Method}} & \multicolumn{4}{c}{\textbf{MMAU Test-mini (1k audios)}} & \multicolumn{4}{c}{\textbf{MMAU Test-full (9k audios)}} \\
             \midrule
              & & Sound & Music & Speech & Avg. & Sound & Music & Speech & Avg. \\
              \hline
              \multicolumn{10}{l}{\textit{\textbf{Baselines:}}} \\
                \color{gray}{-----} & \color{gray}{Human~\cite{sakshi2025mmau}} & \color{gray}{86.3} & \color{gray}{78.2} & \color{gray}{82.2} & \color{gray}{82.2} & - & - & - & - \\
                Audio Flamingo 2 & Direct Inference~\cite{af2} & 61.6 & \bf{74.0} & 30.9 & 55.5 & 65.1 & \bf{72.9} & 40.3 & 59.4 \\
                Phi-4 Multi-Modal & Direct Inference~\cite{phi-4} & 61.0 & 52.9 & 52.8 & 55.6 & - & - & - & - \\
                GPT-4o Voice Model & Direct Inference~\cite{gpt-4o} & 63.4 & 60.8 & 53.2 & 59.1 & - & - & - & - \\
                Qwen2-Audio-7B-Instruct & Direct Inference~\cite{qwen2-audio} & 55.0 & 51.0 & 42.0 & 49.2 & 45.9 & 53.3 & 45.9 & 52.5 \\
                Qwen2-Audio-7B-Instruct & Reproduced~\cite{qwen2-audio} & 61.6 & 54.5 & 42.0 & 52.7 & - & - & - & - \\
                Qwen2-Audio-7B-Instruct & Zero-Shot-CoT~\cite{audio-cot} & 61.9 & 56.3 & 55.3 & 57.8 & - & - & - & - \\
                Qwen2-Audio-7B-Instruct & CoTA~\cite{audio-reasoner} & 60.1 & 64.3 & 60.7 & 61.7 & - & - & - & -  \\
                Qwen2-Audio-7B-Instruct & R1-AQA~\cite{r1-aqa} & 68.8 & 64.3 & 63.7 & 65.6 & \underline{69.8} & 61.4 & \underline{62.7} & \underline{64.4} \\
                Kimi-Audio & Direct Inference~\cite{kimi} & \textbf{73.3} & 61.7 & 60.7 & 65.2 & - & - & - & - \\
                Qwen2.5-Omni-7B & Direct Inference~\cite{qwen-omni} & 67.9 & \underline{69.2} & 59.8 & 65.6 & - & - & - & - \\
                Qwen2.5-Omni-7B & Reproduced & 69.4 & 66.8 & \textbf{61.6} & \underline{65.9} & \bf{71.6} & \underline{67.1} & \bf{66.5} & \bf{68.4} \\
                Qwen2.5-Omni-7B & SARI\textsuperscript{\textdagger}~\cite{sari} & \underline{72.8} & 67.2 & \underline{61.3} & \textbf{67.1} & - & - & - & - \\        
                \hline
                \multicolumn{10}{l}{\textit{\textbf{Ours:}}} \\
                Qwen2.5-Omni-7B & Omni-R1 (AVQA) & 70.9 & 70.1 & \underline{64.9} & 68.6 & 73.6 & 68.6 & \textbf{70.1} & \underline{70.8} \\
                Qwen2.5-Omni-7B & Omni-R1 (AVQA-GPT) & \underline{72.4} & \underline{73.1} & 64.3 & \underline{69.9} & \bf{74.3} & \underline{70.2} & \underline{69.2} & \bf{71.2} \\
                Qwen2.5-Omni-7B & Omni-R1 (VGGS-GPT) & \textbf{73.6} & \textbf{74.3} & \textbf{66.1} & \textbf{71.3} & \underline{74.1} & \bf{70.8} & 68.7 & \bf{71.2} \\
             \hline
        \end{tabular}%
    }
    \label{tab:main_results}
\end{table*}

The learning update for the current policy $\pi_{\theta}$ involves maximizing the GRPO objective function. The expectation is taken over the distribution of input questions $P(Q)$ and the outputs sampled from the old policy for each question:
\begin{equation} \label{eq:GRPO_objective_revised}
\footnotesize 
\begin{split}
    \mathcal{J}_{GRPO}(\theta) &= \mathbb{E}_{\mathcal{D}} \Bigg[\frac{1}{G}\sum_{i=1}^G\frac{1}{|o_i|} \sum_{t=1}^{|o_i|} \Bigg\{ \min \Bigg[ \rho_{i,t} \hat{A}_{i,t}, \\
    & \quad \left. \operatorname{clip} \left( \rho_{i,t}, 1 - \epsilon, 1 + \epsilon \right)  \hat{A}_{i,t} \right] - \beta {\KL}_{i,t}\left[\pi_{\theta} || \pi_{ref}\right]\Bigg\}\Bigg],
\end{split}
\end{equation}
where $\rho_{i,t} = \frac{\pi_\theta(o_{i,t} | q, o_{i,<t})}{\pi_{\theta_{old}}(o_{i,t} | q, o_{i,<t})}$ is the probability ratio between the current policy $\pi_\theta$ and the policy $\pi_{\theta_{old}}$ used for sampling and $\mathcal{D} = [q \sim P(Q), \{o_i\}_{i=1}^G \sim \pi_{\theta_{old}}(O|q)]$ denote the sampling distribution over questions and their corresponding outputs. The $\operatorname{clip}(x, min\_val, max\_val)$ function limits the value of $x$ to be within the range $[min\_val, max\_val]$. The hyperparameter $\epsilon$ defines the PPO clipping range. The per-token KL divergence ${\KL}_{i,t}\left[\pi_{\theta} || \pi_{ref}\right]$ between the current policy $\pi_\theta$ and a reference policy $\pi_{ref}$ is weighted by the hyperparameter $\beta$:
\begin{equation} \label{eq:GRPO_KL_estimator}
\footnotesize 
 {\KL}_{i,t}\left[\pi_{\theta} || \pi_{ref}\right] \approx \frac{\pi_{ref}(o_{i,t}|q,o_{i,<t})}{\pi_{\theta}(o_{i,t}|q,o_{i,<t})} - \log\frac{\pi_{ref}(o_{i,t}|q,o_{i,<t})}{\pi_{\theta}(o_{i,t}|q,o_{i,<t})} - 1.
\end{equation}

\subsection{Automatic Q/A Dataset Generation with ChatGPT}
\label{sec:dataset}
We propose to generate audio question answer datasets automatically.
Given an audio dataset, we first acquire audio-based text captions from an audio LLM.
Specifically, we use the captions generated by Qwen-2 Audio~\cite{qwen2-audio} in the AudioSetCaps dataset~\cite{bai2024audiosetcaps}, which is a compilation of captions generated for many open-source audio datasets.
We use multiple captions which describe the overall acoustic scene, the speech present in the audio, and the music present in the audio.
Next, we use the captions and prompt ChatGPT to generate questions and answers using the provided captions.
We instruct the model to generate a correct answer and three other answers which are similar but clearly incorrect (prompt shown in Figure~\ref{fig:overview}).
We reviewed the generated questions and answers and found the quality to be reasonable, however, we noticed that the original captions feature hallucinations especially about music not present in the audio.
The generated questions could then ask about some music or sounds not actually present in the audio.
We also manually randomized the order of the correct answer within the four choices to ensure a balance.

To verify the potential of such automatically generated training data, we first generated questions for the 40k audios in the AVQA training dataset and name our dataset AVQA-GPT.
Using AVQA-GPT, we are able to compare the performance of our models trained on our automatically curated questions vs the original, ground truth human annotated questions.
Since the two datasets share the same audio files but have different text question and answers, it would verify the quality of our dataset.
In Section~\ref{sec:main_results}, we show that training on our AVQA-GPT dataset can lead to even better performance than training on the AVQA dataset.
Since this approach shows promise, we scaled our method to 182k audio files in the VGGSound training dataset~\cite{chen2020vggsound} and name this dataset VGGS-GPT.
Note that the audio data in AVQA is a subset of the audio data in VGGSound.
Due to GPU memory limitations, we filtered the dataset by length of the question and answer pairs, and only used 54k samples from our VGGS-GPT dataset for training.
As shown in Section~\ref{sec:main_results}, training on this extra data could lead to even better performance.

\begin{table*}[t]
    \renewcommand{\arraystretch}{1.3}
    \small
    \centering
    \caption{Accuracies (\%) on MMAU (v05.15.25). \textsuperscript{\textdagger}Qwen2.5 scores are from \url{https://sakshi113.github.io/mmau_homepage/}}
    \vspace{-2mm}
    \resizebox{0.83\linewidth}{!}{%
        \begin{tabular}{cccccc|cccc}
             \hline
            \toprule
             \multicolumn{1}{c}{\textbf{Model}} & \multicolumn{1}{c}{\textbf{Method}} & \multicolumn{4}{c}{\textbf{MMAU Test-mini (1k audios)}} & \multicolumn{4}{c}{\textbf{MMAU Test-full (9k audios)}} \\
             \midrule
              & & Sound & Music & Speech & Avg. & Sound & Music & Speech & Avg. \\
              \hline
              \multicolumn{10}{l}{\textit{\textbf{Baselines:}}} \\
                Qwen2.5-Omni-7B\textsuperscript{\textdagger} & Direct Inference~\cite{qwen-omni} & - & - & - & - & \underline{74.7} & \textbf{68.6} & \textbf{76.0} & \textbf{73.1} \\
                Qwen2.5-Omni-7B & Reproduced & 78.1 & 65.9 & 70.6 & 71.5 & \textbf{74.9} & \underline{68.0} & \underline{72.4} & \underline{71.8} \\   
                \hline
                \multicolumn{10}{l}{\textit{\textbf{Ours:}}} \\
                Qwen2.5-Omni-7B & Omni-R1 (AVQA) & \underline{82.0} & 70.1 & \underline{75.4} & 75.8 & 77.8 & 69.1 & \textbf{76.3} & 74.4 \\
                Qwen2.5-Omni-7B & Omni-R1 (AVQA-GPT) & \textbf{82.3} & \underline{72.8} & \underline{75.4} & \underline{76.8} & \textbf{78.9} & \textbf{70.9} & \underline{76.0} & \textbf{75.3} \\
                Qwen2.5-Omni-7B & Omni-R1 (VGGS-GPT) & 81.7 & \textbf{73.4} & \textbf{76.0} & \textbf{77.0} & \underline{78.3} & \underline{70.8} & 75.8 & \underline{75.0} \\
             \hline
        \end{tabular}%
    }
    \label{tab:main_results_v2}
\end{table*}
\section{Experiments}

In this section, we report the results on the MMAU~\cite{sakshi2025mmau} and MMAR benchmarks.
There are two versions of MMAU.
The first is the original version, which we denote as MMAU (Original).
All of the previous and concurrent methods have been evaluated on this version.
Very recently, an updated version of the benchmark was proposed\footnote{\url{https://github.com/Sakshi113/MMAU}}, which we denote as MMAU (v05.15.25).
According to the authors, 25\% of the questions and answers have been revised for improved clarity and quality, while 5\% of the audio files have been refined to enhance consistency and fidelity.
We present the results on MMAU (Original) in Section~\ref{sec:main_results} and on MMAU (v05.15.25) in Section~\ref{sec:results-v2}.
We present the results on MMAR in Section~\ref{sec:results-mmar}.

\subsection{Experimental Setup}

To train our models, we use a node with 4 A6000 GPUs (48GB) and 500 GB of RAM.
The batch size per GPU is 1 with gradient accumulation steps of 2 for a total effective batch size of 8.
We train for 1000 steps on AVQA and AVQA-GPT and 2000 steps on VGGS-GPT.
We use a learning rate of $1\times10^{-6}$, temperature of 1.2, 4 responses per GRPO step, and a KL coefficient $\beta$ of 0.04.

\subsection{Main Results on MMAU (Original)}
\label{sec:main_results}
Table~\ref{tab:main_results} shows the main results on MMAU (Original)~\cite{sakshi2025mmau}.
For the baselines, we show the recent methods which achieve State-of-the-Art (SOTA) performance.
We refer readers to MMAU~\cite{sakshi2025mmau} for the older baselines \cite{deshmukh2023pengi, gong2024listen, gong2023joint,deng2024musilingo, liu2024music, liu2023m, ghosh2024gama, qwen2-audio,tangsalmonn, qwenaudio}. 
Note that all of the methods were evaluated on the Test-mini split with 1k audio samples, while only a few of them evaluated on the Test-full split with 9k audio samples.

Qwen2-Audio~\cite{qwen2-audio} was one of the strongest baselines and several methods are based on it.
For example, Zero-Shot-CoT~\cite{audio-cot} and CoTA~\cite{audio-reasoner} try to enhance the model's reasoning capabilities by encouraging the model to output more of the thinking process.
Qwen2.5-Omni~\cite{qwen-omni} is a more recent model which reaches a significantly better average score. 
Although Kimi-Audio~\cite{kimi} and Audio-Flamingo 2~\cite{af2} obtain the best scores on the sound and music categories, Qwen2.5-Omni obtains the best average score of 65.6\% (65.9\% via our reproduction).
SARI~\cite{sari} adapts Qwen2.5-Omni and improves the average score to 67.1\%, which is the current SOTA.
However, they calculate the score using an LLM as a judge instead of the usual method based on string matching with the correct answer, therefore their scores might not be directly comparable with the other results.
On the larger Test-full set of 9k audios, Audio-Flamingo 2 performs the best on music, while Qwen2.5-Omni obtains the highest scores on the sound, speech, and overall average (68.4\%) categories.

Next, we show our Omni-R1 models which fine-tune Qwen2.5-Omni with GRPO.
First, compared to the base model, Omni-R1 fine-tuned on AVQA~\cite{yang2022avqa} improves the average performance on Test-mini from 65.9\% to 68.6\% and on Test-full from 68.4\% to 70.8\%.
Next, we show the result of fine-tuning on our proposed AVQA-GPT dataset.
It uses the same audio as AVQA but the questions are generated by ChatGPT using audio-based text captions from Qwen-2 Audio~\cite{bai2024audiosetcaps}. 
Omni-R1 fine-tuned on AVQA-GPT improves the average performance on Test-mini to 69.9\% and on Test-full from to 71.2\%.
Although AVQA and AVQA-GPT use the same audio files, they use different questions.
It verifies the quality of the questions generated by ChatGPT and shows that they are potentially more useful than human generated questions for fine-tuning.
Finally, we show the result of fine-tuning on our proposed VGGS-GPT, a larger dataset of questions using the same question generation technique as AVQA-GPT, but scaling to more audio from the VGGSound dataset. 
It achieves the best average performance of \textbf{71.3\%} on Test-mini and \textbf{71.2\%} on Test-full.
Omni-R1 also achieves the highest scores on the sound, music, and speech categories comparing all previous models both on Test-mini and Test-full.
It shows that scaling the fine-tuning data can lead to improvements.
Compared to the base model, Omni-R1 improved the average scores by 5.4\% absolute and 2.8\% absolute on Test-mini and Test-full.
Finally, compared to SARI which is the most related method and also builds on the same base model, our average Test-mini scores are 4.2\% absolute higher.

\begin{table*}[t]
\renewcommand{\arraystretch}{1.3}
\small
\centering
\caption{
Accuracies (\%) on MMAR. Mix1: Sound-Music; Mix2: Sound-Speech; Mix3: Music-Speech; Mix4: Sound-Music-Speech.
}
\vspace{-2mm}
\resizebox{0.85\linewidth}{!}{%
\begin{tabular}{cccccccccc}
\toprule
\multirow{2.5}{*}{\textbf{Model}} & \multirow{2.5}{*}{\textbf{Method}} & \multicolumn{3}{c}{\textbf{Single Modality }} &  \multicolumn{4}{c}{\textbf{Mixed Modalities}} & \multirow{2.5}{*}{\textbf{Avg (\%)}} \\ 
\cmidrule{3-9}
&& \textbf{Sound} & \textbf{Music} & \textbf{Speech} & \textbf{Mix1} & \textbf{Mix2} & \textbf{Mix3} & \textbf{Mix4} & \\
\midrule
\multicolumn{10}{l}{\textit{\textbf{Closed-Source Models:}}} \\
Gemini 2.0 Flash  & Direct Inference~\cite{google_gemini_2_flash_2025} & \textbf{61.2} & \textbf{51.0} & \textbf{72.1} & \textbf{81.8} & \textbf{72.5} & \underline{65.9} & \underline{70.8} & \textbf{65.6}   \\ 
GPT-4o Audio & Direct Inference~\cite{gpt-4o-openai} & 53.9 & \textbf{51.0} & \underline{70.4} & \underline{63.6} & \textbf{72.5} & 62.2 & \textbf{75.0} & \underline{63.5} \\
\hline
\multicolumn{10}{l}{\textit{\textbf{Open-Source Models:}}} \\
Qwen2-Audio-7B-Instruct      &    Direct Inference~\cite{qwen2-audio}    & 33.3 & 24.3 & 32.3 & {\ul 9.1} & 31.2 & 30.5 & 25.0 & 30.0  \\
Qwen2.5-Omni-7B & Direct Inference~\cite{qwen-omni} & 58.8 & 40.8 & \textbf{59.9} & \textbf{54.6} & \textbf{61.9} & \textbf{67.1} & 58.3 & {\ul 56.7} \\
Qwen2.5-Omni-7B & Reproduced & \textbf{61.2} & \textbf{47.6} & {\ul 59.5} & \textbf{54.6} & {\ul 61.5} & {\ul 61.0} & \textbf{66.7} & \textbf{58.0} \\
\hline
\multicolumn{10}{l}{\textit{\textbf{Ours:}}} \\
Qwen2.5-Omni-7B & Omni-R1 (AVQA) &  {\ul 63.6} & {\ul 51.5} & 62.2 & \textbf{72.7} & 65.1 & 62.2 & \textbf{70.8} & 61.2 \\
Qwen2.5-Omni-7B & Omni-R1 (AVQA-GPT) & 58.8 & \textbf{53.4} & \textbf{65.0} & {\ul 45.5} & {\ul 66.1} & {\ul 63.4} & {\ul 66.7} & {\ul 61.5} \\
Qwen2.5-Omni-7B & Omni-R1 (VGGS-GPT) & \textbf{67.3} & {\ul 51.5} & {\ul 64.3} & {\ul 45.5} & \textbf{70.2} & \textbf{64.6} & \textbf{70.8} & \textbf{63.4} \\

\bottomrule
\end{tabular}%
}
\label{tab:main_mmar}
\end{table*}

\subsection{Main Results on MMAU (v05.15.25)}
\label{sec:results-v2}

Table~\ref{tab:main_results_v2} shows the results on MMAU (v05.15.25), the updated version of the benchmark. 
We focus on Qwen2.5-Omni-7B since it achieves the highest scores on this version.
The authors of MMAU reported the scores of Qwen2.5-Omni-7B on the benchmark website with the Test-full split but not with the Test-mini split.
We evaluated Qwen2.5-Omni-7B on both splits and the model achieved average scores of 71.5\% on Test-mini and 71.8\% on Test-full.
Comparing our reproduced results vs the results presented on the benchmark website using Test-full, our reproduced results are slightly lower (71.8\% reproduced vs 73.1\% reported).
Since the Test-full answers are hidden and the scores are computed by an online server, we have reached out to the authors to inquire why our reproduced results are slightly worse than theirs.

Next, we show our Omni-R1 models which fine-tune Qwen2.5-Omni with GRPO.
Similar to the results on MMAU (Original), Omni-R1 achieves significant improvements over the base model on MMAU (v05.15.25).
First, fine-tuning on AVQA results in average scores of 75.8\% and 74.4\% on Test-mini and Test-full respectively.
This shows the effectiveness of the GRPO fine-tuning process.
Second, fine-tuning on AVQA-GPT results in average scores of 76.8\% and 75.3\% on Test-mini and Test-full respectively.
Fine-tuning on AVQA-GPT results in better scores than fine-tuning on AVQA, which shows the effectiveness of our automatically generated question and answers.
The average score of \textbf{75.3\%} is our best result on Test-Full and it outperforms Qwen2.5-Omni-7B's average score of 71.8\% on Test-full by 3.5\% absolute.
Finally, fine-tuning on AVQA-GPT results in average scores of 77.0\% and 75.0\% on Test-mini and Test-full respectively.
The average score of \textbf{77.0\%} is our best result on Test-mini and it outperforms Qwen2.5-Omni-7B's average score of 71.5\% on Test-full by 5.5\% absolute.
Overall these results show that our fine-tuning process is robust and shows consistent gains compared to the baseline method even with updates to the benchmark.

\subsection{Main Results on MMAR}
\label{sec:results-mmar}
Table~\ref{tab:main_mmar} shows the results on MMAR~\cite{ma2025mmar}, which only provides a test set of 1k audios.
For the baselines, we show the results of Qwen2-Audio and Qwen-2.5-Omni and refer the reader to the original paper for the other baselines.
Our reproduction of the base Qwen-2.5-Omni results in an average score of 58\%.

Next, we show our Omni-R1 models which fine-tune Qwen2.5-Omni with GRPO.
First, Omni-R1 fine-tuned on AVQA achieves an average score of 61.2\% (3.2\% absolute higher than the base model), showing the effectiveness of our fine-tuning method.
Next, fine-tuning on our automtically generated AVQA-GPT and VGGS-GPT datasets results in further gains.
Our best model fine-tuned on VGGS-GPT achieves an average score of 63.4\% (5.4\% absolute higher than the base model), setting a new State-of-the-Art for open-source models.
Our fine-tuned models improved the scores on all categories compared to the base model, except on mixtures of sound and music (Mix1).
However, this subset only contains 11 test samples in total and the number of correct answers changed from 6 to 5 which is acceptable.
Our model's performance approaches the scores of strong closed-source models such as Gemini and GPT-4o, shown at the top of the table.
Finally, we note that while the MMAR dataset was designed to elicit deep reasoning from audio LLMs, our model could achieve new State-of-the-Art results without outputting long text reasoning steps and simply outputting the correct answer.

\subsection{Fine-Tuning Audio LLMs without Audio}
In this section, we investigate how the models improve with GRPO fine-tuning.
In particular, we wanted to understand why GRPO improved Qwen2.5-Omni's performance less than Qwen2.5-Omni's on MMAU (Original).
Comparing the result of fine-tuning Qwen2.5-Omni with GRPO (Omni-R1, ours) and Qwen2-Audio with GRPO (R1-AQA\cite{r1-aqa}) on AVQA in Table~\ref{tab:main_results}, Qwen2.5-Omni's average score improved from 65.9\% to 68.6\% (2.7\% absolute) while Qwen2-Audio's average score improved from 52.7\% to 65.6\% (12.9\% absolute).
To investigate this, we tested the models without audio to understand their text reasoning capabilities.
We then fine-tune the models without audio (just text) and measure the resulting performance.
Note that we use MMAU (Original) for these experiments.

Table~\ref{tab:ablation_audio} shows the result of performing inference both with audio and without audio.
The scores reported are the averages on MMAU Test-mini split.
For inference without audio, we simply drop the audio tokens from the input.
Without access to the input audio, Qwen2-Audio and Qwen2.5-Omni achieve 30.5\% and 49.3\% respectively.
While the performance for Qwen2-Audio is around chance (25\%), it is surprising that Qwen2.5-Omni performs so well without audio.
It shows that many of the MMAU questions can be answered without audio and that Qwen2.5-Omni has good text-based knowledge about audio.
We suspected that GRPO helped Qwen2-Audio more significantly because its text-based reasoning was worse than Qwen2.5-Omni's.
Therefore, we checked the performance after fine-tuning both models with GRPO on AVQA.
Qwen2-Audio's performance without audio increased significantly from 30.5\% to 44.6\%, while Qwen2.5-Omni's performance without audio increased only slightly from 49.3\% to 51.7\%.
This shows that fine-tuning with GRPO significantly improved Qwen2-Audio's text-based reasoning, but not Qwen2.5-Omni since the improvement was smaller.
This is reasonable since Qwen2.5-Omni's performance without audio was already high.

\begin{table}[t]
    \renewcommand{\arraystretch}{1.2}
    \small
    \centering
    \caption{Ablation of RL fine-tuning with vs.\ without audio across datasets. Accuracies (\%) on MMAU mini at inference time.}
    \vspace{-2mm}
    \label{tab:ablation_audio}
    \resizebox{\columnwidth}{!}{%
    \begin{tabular}{l c c c c}
        \toprule
        \multirow{2}{*}{\textbf{Model}}
      & \textbf{RL FT:}
      & \multirow{2}{*}{\textbf{FT Dataset}}
      & \multicolumn{2}{c}{\textbf{Inference}} \\
        \cmidrule(lr){4-5}
        & \textbf{w/ audio?}& & \textbf{w/ audio} & \textbf{w/o audio} \\
        \midrule
        Qwen2-Audio\ & --           & --       & 52.7 & 30.5 \\
        Qwen2-Audio\ & \cmark & AVQA     & \textbf{63.2} & \textbf{44.6} \\
        Qwen2-Audio\ & \xmark    & AVQA     & 58.8 & \underline{42.4} \\
        Qwen2-Audio\ & \xmark     & ARC–Easy & \underline{60.2} & 42.2 \\

        \midrule
        Qwen2.5-Omni       & --           & --       & 65.9 & \underline{49.3} \\
        Qwen2.5-Omni       & \cmark & AVQA     & \textbf{68.6} & \textbf{51.7} \\
        Qwen2.5-Omni       & \xmark     & AVQA     & 65.6 & 49.2 \\
        Qwen2.5-Omni       & \xmark     & ARC–Easy & \underline{68.2} & \bf{51.7} \\
        \bottomrule
    \end{tabular}}
\end{table}

Since fine-tuning the models with GRPO could improve the performance for inference both with and without audio, we wondered how well it work if we fine-tuned the models \textit{without audio} on AVQA.
Therefore, we fine-tuned the models on AVQA with GRPO to answer the questions just based on the provided text questions and answers.
Despite fine-tuning without audio, this strategy could significantly improve the performance for Qwen2-Audio.
The performance without audio improved from 30.5\% to 42.4\%.
Surprisingly, the performance for inference with audio improved from 52.7\% to 58.8\%.
This is surprising because we expected that fine-tuning on just text would make the performance worse as it could lead to catastrophic forgetting with respect to handling the audio input.
This actually happened for Qwen2.5-Omni: the performance slightly decreased compared to the base model.
Overall, these results are consistent with our hypothesis that much of Qwen2-Audio's improvement with GRPO could be attributed to improving the text-based reasoning.
Meanwhile, Qwen2.5-Omni's stronger base text knowledge leads to smaller improvements with GRPO.
Note that fine-tuning \textit{with audio} on AVQA still performed better both for inference with and without audio.

Since fine-tuning the models without audio on AVQA could improve the base model performance, we decided to fine-tune on a text-based Q/A dataset without audio.
Therefore, we fine-tuned on the ARC-Easy dataset~\cite{clark2018think}, which contains Q/A questions about science in the same format as MMAU (one question and four answer choices). 
Interestingly, fine-tuning on this dataset resulted in even better performance than fine-tuning on AVQA without audio.
Notably, Qwen2.5-Omni's performance with audio after training on text-only ARC-Easy (68.2\%) nearly matches the performance of training on AVQA with audio (68.6\%).
However, the performance was not as good as fine-tuning on AVQA with audio.
This provides further evidence that the GRPO fine-tuning is mainly improving the text-based reasoning of the models.

Overall, our results show: 
1.) Qwen2.5-Omni is a much stronger text-based reasoner than Qwen2-Audio 
2.) Most of the improvement from fine-tuning Qwen2-Audio with GRPO can be attributed to the improvement in text-based reasoning 
3.) Fine-tuning Qwen2.5-Omni with GRPO results in a smaller performance boost since the base text-based reasoning is already strong
4.) Fine-tuning the models on text-only Q/A datasets is surprisingly effective, but fine-tuning on Q/A datasets with audio still works better.

\section{Conclusion}
We propose Omni-R1, an Audio LLM based on fine-tuning the multi-modal LLM Qwen2.5-Omni with GRPO for better audio question answering. 
We use a simple yet effective prompt which is straight to the point and makes training and testing efficient.
On the MMAU benchmark, Omni-R1 sets new State-of-the-Art results across sounds, music, speech, and overall accuracy on both Test-mini and Test-full. 
On the MMAR benchmark, our method achieves new State-of-the-Art results among open-source models and competitive performance compared to significantly larger, closed-source models.
To scale up, we also introduced two large, auto-generated audio question answering datasets (AVQA-GPT and VGGS-GPT), which further boost performance. 
Through our experiments of testing the models with and without audio, we showed that much of the gains come from better text-based reasoning. 
Surprisingly, fine-tuning with only text (no audio) also delivers strong improvements in audio question answering. 

On the one hand, since multi-modal LLMs usually start with a text-only LLM and add multi-modal encoders, it's reasonable that fine-tuning Audio LLMs on text-only datasets and improving the base knowledge would be helpful for the multi-modal abilities.
On the other hand, it's surprising that fine-tuning on a text-only dataset could help the audio-based performance so much.
This suggests future work is necessary on better text-only datasets to help multi-modal LLMs and on curriculum training of text-only and audio-text datasets.

Finally, collecting transcribed audio or human-annotated audio can be expensive.
Our findings suggest that Audio LLMs can be steered towards better capabilities using text-only data.
Moreoever, our best models were fine-tuned on audio data with questions automatically generated without human annotators.
These results could help reduce the costs of building intelligent audio-based agents.

\section{Acknowledgments}
This research was supported by the MIT-IBM Watson AI Lab and an NDSEG Fellowship to A.R.

\bibliographystyle{IEEEtran}
\bibliography{bib}

\end{document}